\documentclass[bst/sn-mathphys,iicol]{sn-jnl}

\usepackage{textcomp}
\usepackage{cleveref}
\crefname{section}{§}{§§}

\usepackage{graphicx}%
\usepackage{multirow}%
\usepackage{amsmath,amssymb,amsfonts,epsf}%
\usepackage{amsthm}%
\usepackage{mathrsfs}%
\usepackage[title]{appendix}%
\usepackage{xcolor}%
\usepackage{textcomp}%
\usepackage{manyfoot}%

\usepackage{color}
\usepackage{varwidth}

\DeclareMathOperator*{\argmin}{arg\,min}
\DeclareMathOperator*{\maximize}{maximize}
\usepackage{algorithmicx}
\usepackage{algorithm}

\usepackage[nolist]{acronym}
\begin{acronym}
	\acro{AWGN}[AWGN]{Additive White Gaussian Noise}
	\acro{LoS}[LoS]{Line of Sight}
	\acro{RF}[RF]{Radio Frequency}
	\acro{SNR}[SNR]{Signal-to-Noise Ratio}
	\acro{UAV}[UAV]{Unmanned Aerial Vehicle}
	\acro{OCC}[OCC]{Optical Camera Communications}
	\acro{UV}[UV]{Ultraviolet}
    \acro{GNSS}[GNSS]{Global Navigation Satellite}
    \acro{UWB}[UWB]{Ultra Wide-band}
    \acro{NRZ}[NRZ]{Non-Return-to-Zero}
    \acro{OOK}[OOK]{On-Off Keying} 
    \acro{FoV}[FoV]{Field of View}
    \acro{IPI}[IPI]{Inter Pixel Interference}
    \acro{SISO}[SISO]{Single Input Single Output}
    \acro{BER}[BER]{Bit Error Rate}
    \acro{FFT}[FFT]{Fast Fourier Transform}
  \acro{UVDAR}[\rm{UVDAR}]{UltraViolet Direction And Ranging}
  \acro{IR}[IR]{Infrared}
\end{acronym}

\usepackage{siunitx}

\jyear{2022}%

\numberwithin{equation}{section}

\theoremstyle{thmstyleone}%
\theoremstyle{thmstyletwo}%

\theoremstyle{thmstylethree}%

\newcommand{\PREPRINTYEAR}{2025}
\newcommand{\PUBLISHEDIN}{Autonomous Robots}

\begin{document}
\thispagestyle{empty}
{
  \topskip0pt
  \vspace*{\fill}
  \centering
  \LARGE{%
    \copyright{} \PREPRINTYEAR~\PUBLISHEDIN\\\vspace{1cm}
    Personal use of this material is permitted.
    Permission from \PUBLISHEDIN~must be obtained for all other uses, in any current or future media, including reprinting or republishing this material for advertising or promotional purposes, creating new collective works, for resale or redistribution to servers or lists, or reuse of any copyrighted component of this work in other works.}
    \vspace*{\fill}
}

\title[Optical communication-based identification for multi-UAV systems: theory and practice]{Optical communication-based identification for multi-UAV systems: theory and practice}


\author*[1,2]{\fnm{Daniel} \sur{Bonilla Licea}}\email{daniel.bonilla@um6p.ma}
\author[2]{\fnm{Viktor} \sur{Walter}}\email{viktor.walter@fel.cvut.cz}
\author[1]{\fnm{Mounir} \sur{Ghogho}}\email{mounir.ghogho@um6p.ma}
\author[2]{\fnm{Martin} \sur{Saska}}\email{martin.saska@fel.cvut.cz}
\affil[1]{\orgname{Mohammed VI Polytechnic University}, \orgaddress{\country{Morocco}}}
\affil[2]{\orgname{Czech Technical University in Prague}, \orgaddress{\country{Czechia}}}


\abstract{Mutual relative localization and identification are important features for multi-\ac{UAV} systems. Camera-based communications technology, also known as \ac{OCC} in the literature, is a novel technology that brings a valuable solution to this task. In such a system, the \acp{UAV} are equipped with LEDs acting as beacons, and with cameras to locate the LEDs of the other \acp{UAV}. Specific blinking sequences are assigned to the LEDs of each of the \acp{UAV} to uniquely identify them. This camera-based system is immune to \ac{RF} electromagnetic interference and operates in \ac{GNSS}-denied environments. In addition, the implementation of this system is inexpensive. In this article, we study in detail the capacity of this system and its limitations. Furthermore, we show how to construct blinking sequences for \ac{UAV} LEDs to improve system performance. Finally, experimental results are presented to corroborate the analytical derivations. }


\maketitle
\section{Introduction}
\label{sec:intro}
Mutual relative localization and identification are important features in multi-\ac{UAV}. While relative localization is important for close cooperative flying and mutual collision avoidance, the identification of neighboring team members is crucial for high-level planning. This feature can be implemented using \ac{RF} electromagnetic signals, vision-based techniques, or through a combination of both. For instance, in \cite{Ali2014ARO,Vilca2016}, RTK-\ac{GNSS} is used for \ac{UAV} localization. In \cite{ChenCCDC2018}, \ac{UWB} ranging is used to determine the distance between the \acp{UAV}. In \cite{MercadoECC2013}, they use a motion capture system that sends its position estimate to the \ac{UAV} via an \ac{RF} link. 
Such localization techniques based on \ac{RF} are vulnerable to electromagnetic interference and may fail in providing mutual identification in multi-robot systems. On the other hand, vision-based techniques are immune to \ac{RF} electromagnetic interference and can provide relative localization and identification in multi-robot systems. Although vision-based systems are vulnerable to optical attacks, see section \ref{sec:discussion:attack}.
\par
Vision-based localization and identification systems can be divided into passive and active, distinguished based on whether the optical markers emit light. In \cite{Krajnik2014JINT}, the authors present a passive system where specific marker patterns are assigned to each robot. The same system was used in \cite{Saska2017ARo} for outdoors localization and identification of \acp{UAV}. A disadvantage of passive systems is their sensitivity to ambient light, making them ineffective in poorly illuminated environments. This is solved by using active systems where the markers of \acp{UAV} are generally implemented with LEDs.
\par
In \cite{LEDCAM1}, the authors equipped a \ac{UAV} with infrared LEDs and used a CMOS camera to perform indoors localization. Different blinking frequencies, in the range of 1-2kHz, were assigned to each LED to differentiate them. The blinking frequencies were set in such a way that no two signals shared common harmonics to avoid ambiguities in their discrimination. In \cite{UVDAR2}, our research group presented the \ac{UVDAR} system for \acp{UAV}, see Fig. \ref{fig:uav_platform}. In this system, \acp{UAV} are equipped with \ac{UV} LEDs as markers and cameras coupled with optical \ac{UV} bandpass filters. The optical signals emitted by the LEDs were square signals of different frequencies. This is a simple way to discriminate the different blinking patterns, but it is inefficient as will be shown in this paper.
\par
In the field of communications, the \acf{OCC} system, in which the transmitter is a LED and the receiver is a camera \cite{LE201795,SAEED2019100900}, has been used for car-to-car communications and car-to-infrastructure communications \cite{NguyenIEEEPJ2017,SoaresCSNDSP2020}, and recently it has started to being applied to communications with \acp{UAV} \cite{TakanoVTC2021,TakanoIEEEA2022}. Despite the fact that the purpose of the \ac{OCC} system is to exchange information, 
it can be used to improve the active vision-based localization and identification systems for \acp{UAV}.  
\par
This paper focuses on the mutual identification capacity of a vision-based active system, as the relative localization issue is beyond its scope.
We investigate the mutual identification using the \ac{UVDAR} system, as mentioned above.
\par
\par
\par
The main contributions of this article are as follows:
\begin{itemize}
\item Blinking sequence generation method: a theoretical framework was developed to design sets of blinking sequences for the LEDs of \ac{UAV} groups. These sequence sets are optimized to discriminate between as many sequences as possible in the shortest time. This enables large groups of \acp{UAV} to perform mutual identification in the shortest time.
\item Theoretical analysis of \ac{UVDAR}: we performed an analytical analysis to derive the probability of misdetection of the blinking sequences, and analytically determined the number of different blinking sequences that can be detected as a function of their length. This can be used to calculate the total number of \acp{UAV} that can be identified by this system.
\item Experimental validation: we implemented a prototype of the proposed vision-based mutual identification system for \acp{UAV} and tested it outdoors.
\end{itemize}
The paper is organized as follows. In section \ref{sec:syst}, we describe the system model. This includes the models for the \ac{UAV} clock signal, for the optical identification system, and for the optical transmission channel. In section \ref{sec:Problem}, we formalize the problem of the visual identification system for \acp{UAV}. In section \ref{sec:binarySeq}, we describe the theoretical framework for the construction of the blinking sequences for the LEDs using the \ac{NRZ} optical modulation (a popular optical modulation scheme). In section \ref{sec:detection:Manchester}, we show how, instead of the \ac{NRZ}, the use of the Manchester optical modulation (another popular optical modulation scheme) changes the properties of the identification system. In section \ref{sec:clock}, we study the effects of the \acp{UAV} clock on the performance of the identification system. Section \ref{sec:experiments} describes the experiments performed. Section \ref{sec:discussion} discusses some additional aspects of the system. Conclusions are provided in section \ref{sec:conc}. 

\section{System Model}
\label{sec:syst}
\begin{figure}[tp]
    \centering
      \includegraphics[trim={0 10.5cm 0 0.3cm},clip,width=0.45\textwidth]{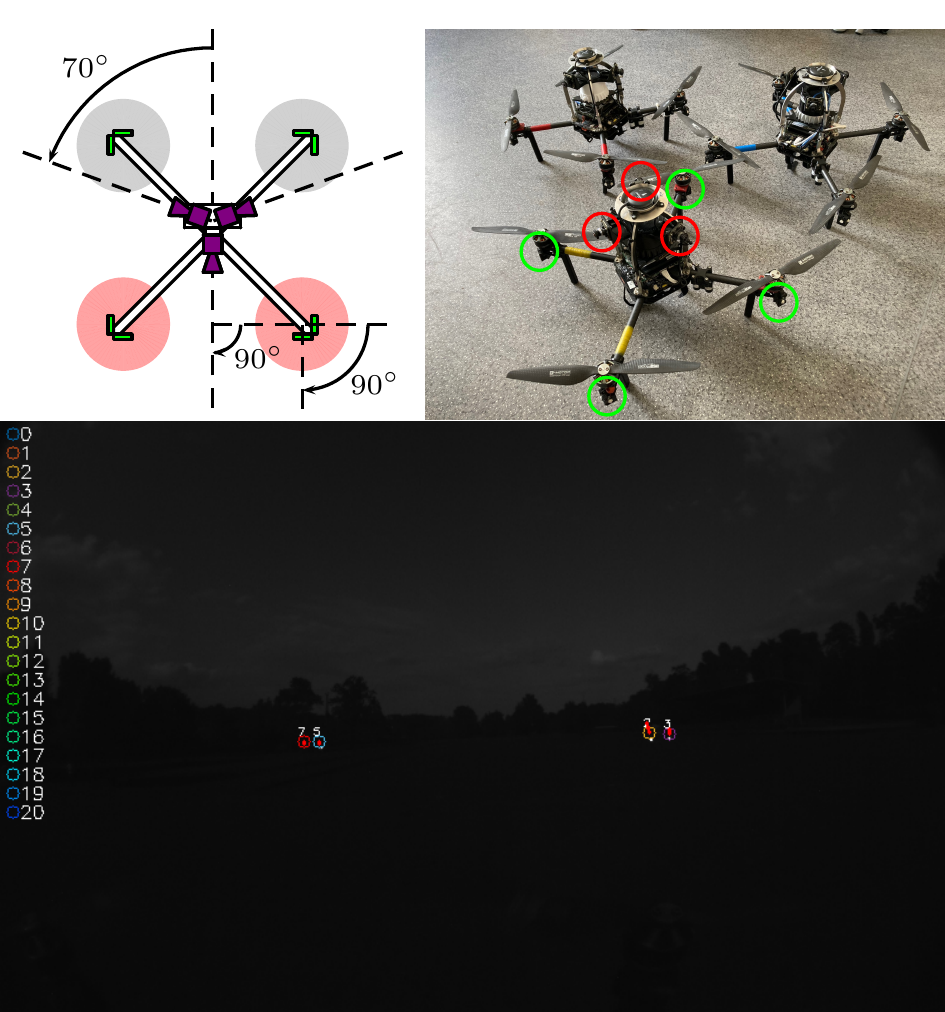}
    \caption{On top \ac{UAV} platforms used in the experimental data acquisition.
    The frame is based on the \emph{Holybro X500} platform, with an arm length from the center of \SI{0.245}{\centi\meter}.
    Each unit is equipped with the \ac{UVDAR} system, with three \ac{UV} cameras and four pairs of \ac{UV} LEDs.
    Each of the LED pairs is placed at the end of each arm, and cameras are attached as shown on the diagram to cover the entire horizontal surroundings.
      The LEDs are rated at \SI{1}{\watt} input power, but we are driving them at {\SI{600}{\milli\watt}}, producing cca {\SI{276}{\milli\watt}} of radiometric power.
    On the bottom, view of the left \ac{UVDAR} camera of \ac{UAV}-0 in the outdoor flight experiment.
      The markers are correctly labeled based on the retrieved signal.
    }
      \label{fig:uvdar_view_experiment}
      \label{fig:3_uav_platforms}%
      \label{fig:uav_platform}%
    \end{figure}

We consider a group of \acp{UAV} composed of $J$ members. Each \ac{UAV} is equipped with the \ac{UVDAR} system, see Fig. \ref{fig:uav_platform}, which is composed mainly of three modules, see Fig. \ref{fig:diagrams}: the optical transmitter, the optical receiver, and the clock signal generator. We now discuss the three modules, and the optical channel model.
\subsection{Clock signal}
\label{sec:syst:clock}
The clock signal's falling (or rising) edges indicate the instants when the receiver's camera shoots and when the LEDs of the transmitter can change state. An ideal clock signal should be stable and the interval between falling (or rising) edges should always remain constant. Furthermore, the clock signal frequency of the different \acp{UAV} of the group should be exactly the same. Unfortunately, due to physical impairments, the clock signals are not perfectly stable. Even if the nominal frequencies of all the clocks are the same, their true frequencies will differ slightly. These impairments and their effects on similar systems have been documented in the literature. For instance, in \cite{NguyenICUFN2014,HuMobiCom13}, it was observed that the measured interframe interval of certain cameras is time-variant. In \cite{LE201795}, it was noted in the context of smartphone cameras, that the nominal frame rate by software differs from the true frame rate and varies depending on the phone. As will be demonstrated in section \ref{sec:clock}, these irregularities on the clock signal limit the capacity of the optical identification system studied in this article. First, let us describe the model of the $j$th \ac{UAV} clock signal, denoted by $c_j(t)$. Without loss of generality, we consider that the optical transmitter and receiver are controlled by the falling edges of $c_j(t)$. Based on the mathematical models for clock signals described in \cite{WeiIEEETComm,MesserschmittIEEEJsac1990}, we model the $k$th falling edge instant of $c_j(t)$ as:
    \begin{equation}
    \label{eq:clock}
    t_{j,k}=T_j+n_{j,k}+t_{j,k-1},\quad\quad k=1,2,
    \end{equation}
where $t_{j,0}$ is the instant when the system of the $j$th \ac{UAV} is turned on; $t_{j,k}$ with $k\geq1$ is the instant of the $k$th falling edge instant of $c_j(t)$; $T_j>0$ is the true clock signal period of $c_j(t)$ and is modelled as a random variable with mean $\mathbb{E}[T_j]=T$, with $T$ being the nominal period of the clock signal, with variance $\mathrm{var}[T_j]=\sigma^2_T$. Due to the fabrication process uncertainties, different clocks will have slightly different oscillation frequencies, even if their nominal frequencies are the same. Thus, we consider that the set $\{T_{j}\}_{j=1}^J$ is composed of $J$ statistically independent and identically distributed random variables. $n_{j,k}$ accounts for the frequency instability of the clock signal. 

\subsection{Optical Transmitter}
\label{sec:syst:tx}
\begin{figure*}
    \includegraphics[trim={0 23.5cm 0 0},clip,width=1.0\linewidth]{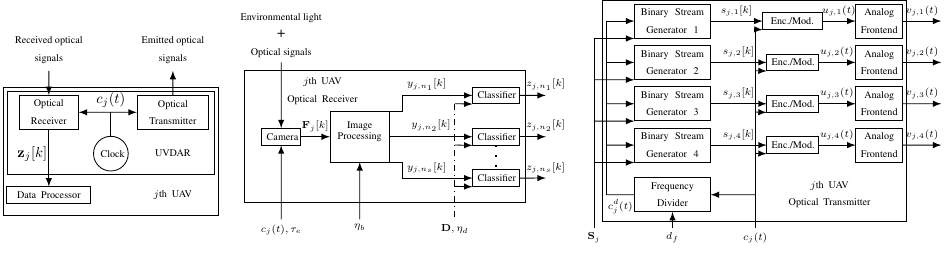}
    \caption{UVDAR block diagram of the $j$th \ac{UAV} (left). Optical receiver architecture for the $j$th UAV (center). Optical transmitter architecture for the $j$th UAV (right). }
    \label{fig:diagrams}
\end{figure*}
Fig. \ref{fig:diagrams} (right)  depicts the diagram of the optical transmitter for the quadrotor shown in Fig. \ref{fig:uav_platform}. The transmitter is divided into $M$ parallel branches. In this particular case, we select $M=4$, i.e., one branch per \ac{UAV} arm in Fig. \ref{fig:uav_platform}. We can have more branches \cite{UVDAR3}, a detailed discussion on such configurations is beyond the scope of this article, but in section \ref{sec:discussion:txdesign} we briefly discuss the implications of the selection of the number of branches $M$. The transmitter modules are the following: {\bf 1) Binary stream generator}: takes as inputs the clock signal $c_j^d(t)$ and a binary matrix $\mathbf{S}_j$ of size $M\times L$, which contains $M$ binary sequences of length $L$. The $m$th binary stream generator produces the discrete-time stream $s_{j,m}$ composed of a continuously repeated concatenation of the binary sequence contained in the $m$th row of the matrix $\mathbf{S}_j$; $s_{j,m}[k]$ denotes the $k$th bit in the binary stream $s_{j,m}$. All the binary sequences used by the \acp{UAV} in the group are stored in a dictionary matrix $\mathbf{D}$; each row of the matrix $\mathbf{D}$ is a different binary sequence, and its identification number is the row number in which it is stored. The dictionary $\mathbf{D}$ is shared by all the \acp{UAV} in the group. Finally, the identification number embedded in the stream $s_{j,m}$ is the same as the identification number of the binary sequence used to generate it. {\bf 2) Encoder and Modulator (Enc./Mod.)}: the encoder codifies the binary discrete-time stream $s_{j,m}$ with a line code, such as \ac{NRZ} or Manchester \cite{ChowIEEEPhotonics2013}. The modulator modulates the encoded signal with \ac{OOK} \cite{KahnPIEEE1997} to produce the continuous-time electrical signal $u_{j,m}(t)\in\left\{0,1\right\}$. {\bf 3) Frequency divider}: divides the clock signal frequency by a factor $d_f$. The modulator and the encoder are both driven directly with the clock signal $c_j(t)$. But, the binary stream generator is driven by $c_j^d(t)$. If the Manchester code is used, then each bit consists of two minibits and the frequency of $c_j^d(t)$ must be half the frequency of $c_j(t)$, i.e., we need $d_f=2$. Alternatively, if the \ac{NRZ} code is used, then each bit consists of only one mini-bit. Thus, the frequencies of $c_j^d(t)$ and $c_j(t)$ must be equal, i.e. $d_f=1$. {\bf 4) Analog frontend}: this transforms the binary electrical signal $u_{j,m}(t)$ into the optical signal $v_{j,m}(t)=Pu_{j,m}(t)$, where $P$ is the emitted optical power. Each frontend has two \ac{UV} LEDs that emit the same optical signal with a wavelength of 395 nm. Both \ac{UV} LEDs are mounted orthogonally (see Fig. \ref{fig:uav_platform}) to increase the angular visibility range of the optical signal.

\subsection{Optical Receiver}
\label{sec:syst:rx}

The architecture of the optical receiver of the $j$th \ac{UAV} is shown in Fig. \ref{fig:diagrams} and is composed of the following modules: {\bf 1) Camera}: a grey scale \ac{UV} sensitive camera mounted on the \ac{UAV} as shown in Fig. \ref{fig:uav_platform} (center). The camera is coupled with an optical filter that allows \ac{UV} light to pass and filters-out most visible light, see \cite{UVfilter}. The filter attenuates most of the background light and facilitates the detection of the \ac{UV} light emitted by the other \acp{UAV}. The camera shoots at every falling edge of the clock signal $c_j(t)$ with an exposure time $\tau_e$. The $k$th frame captured is denoted as $\mathbf{F}_j[k]$.  
{\bf 2) Image Processing}: this module must detect the bright spots potentially generated by the \ac{UV} LEDs from the other \acp{UAV} in the group, track their motion on the screen, and then extract the optical signal from their blinking patterns. To do this, the frame $\mathbf{F}_j[k]$ is first binarized with the threshold $\eta_b$ to produce $\bar{\mathbf{F}}_{j}[k]$. This simplifies the distinction between the background and bright spots potentially generated by \ac{UV} LEDs, see Fig. \ref{fig:uvdar_view_experiment}. When a new bright spot is detected in $\bar{\mathbf{F}}_{j}[k]$, a serialized service number $n_s$ is assigned and the following operations take place simultaneously: i) the coordinates of the central pixel of the $n_s$th bright spot is estimated $\hat{\mathbf{p}}_{j,n_s}[k]$, and its onscreen motion begins to be tracked; ii) the pixel with coordinates $\hat{\mathbf{p}}_{j,n_s}[k]$ is read in $\bar{\mathbf{F}}_{j}[k]$, and the values are stored as a binary time series $y_{j,n_s}[k]$. The instant when the time series associated with the $n_s$ bright spot is created is denoted as its {\it birth time} $t_{b,j,n_s}$; iii) a classifier instance is created to process the time series $y_{j,n_s}$.
As long as the $n_s$th bright spot is successfully tracked, the associated time series $y_{j,n_s}$ remains {\it alive}. But, once the tracking fails, the time series $y_{j,n_s}$ {\it dies} and the associated classifier instance is destroyed. We denote this instant as the {\it death time} $t_{d,j,n_s}$ of time series $y_{j,n_s}$. Possible reasons for tracking failure may include LED occlusions, fast movements of the bright spot on the camera frame, or LED blinking patterns with long times off. {\bf 3) Classifier}. Each classifier takes the dictionary $\mathbf{D}$ described in section \ref{sec:syst:tx} and the last $L$ bits received in the binary stream $y_{j,n_s}$ as input. The classifier output is the time-series $z_{j,n_s}[k]$, which is then fed into the higher level modules. The first objective of the classifier is to determine if $\{y_{j,n_s}[m]\}_{m=k-L+1}^k$ was generated with a binary sequence contained in the dictionary $\mathbf{D}$. This allows for discarding the bright spots generated by sources other than the \acp{UAV}. This is done by calculating the correlation of $\{y_{j,n_s}[m]\}_{m=k-L+1}^k$ with the sequences contained into the dictionary $\mathbf{D}$, and then comparing it with a detection threshold $\eta_d$. If the classifier decides that $\{y_{j,n_s}[m]\}_{m=k-L+1}^k$ was not generated by a binary sequence contained in the dictionary $\mathbf{D}$, then it produces $z_{j,n_s}[k]$$=$$-1$. In the contrary case, the classifier estimates the identification number of the binary stream $y_{j,n_s}$; $z_{j,n_s}[k]$ takes on this number.  

\subsection{Optical channel model}
\label{sec:syst:channel}
We reasonably assume that the exposure time $\tau_e$ is smaller than the coherence time of the background illumination signal. Then, regarding the optical channel between an LED from the $j$th \ac{UAV} and the camera from the $\ell$th \ac{UAV}, we have:
\begin{equation}
    \label{eq:snr:1}
    x_{\ell}[k]=h_{\ell}(t_k)\int_{t_{\ell,k}}^{t_{\ell,k}+\tau_e}v_j(t)\mathrm{d}t+n_\ell[k],
\end{equation}
where $x_{\ell}[k]$ is the pixel value from the $k$th frame captured by the $\ell$th \ac{UAV} camera, $h_{\ell}(t_k)$ is the optical channel gain, $v_j(t)$ is the optical power emitted by the LED of the $j$th \ac{UAV}, see \ref{sec:syst:tx}, and $n_\ell[k]$ is the noise generated at the pixel. In general, the integral representing the exposure process in \eqref{eq:snr:1} becomes \cite{WeiIEEETComm}:
\begin{equation}
\label{eq:snr:2}
    \int_{t_k}^{t_k+\tau_e}{p}(t)\mathrm{d}t=\tau_eP(a[k]s[k_t]+(1-a[k])s[k_t+1]),
\end{equation}
where $k_t$ is related to $k$ by:
\begin{equation}
\label{eq:snr:2b}
    k_t=\left\{\argmin_n \{\mid t_{\ell,k}-t_{j,n}\mid\}:    t_{j,n} \leq t_{\ell,k}<t_{j,n+1}\right\}.
\end{equation}
During the exposure process, a bit transition may occur in $v_j(t)$. This is modelled by the random process $a[k]\in[0,1]$, whose behaviour depends on the relative uncertainties of the clock signals from the transmitter and receiver, as well as on the exposure time $\tau_e$. 

\section{Problem description and proposed solution}
\label{sec:Problem}
The mutual identification system must determine the identification numbers associated with the optical signals emitted by the LEDs of the \acp{UAV} in the group, as in section \ref{sec:syst:tx}. This identification system can be used to estimate the relative location and pose of the \acp{UAV} \cite{UVDAR3}. In this case, each branch of the optical transmitters of the \acp{UAV} will transmit different optical signals, and thus all rows in $\mathbf{S}_j$ will be different. On the other hand, the identification system can be used to estimate only the relative positions of the \acp{UAV} \cite{UVDAR1,UVDAR2,UVDAR4}. In this case, each branch of the optical transmitters will transmit the same optical signals, and therefore all the rows in $\mathbf{S}_j$ will be identical. The matrix $\mathbf{S}_j$ is related to the dictionary $\mathbf{D}$ by:
\begin{equation}
    \label{eq:00S}
    \mathbf{S}_j=\mathbf{A}_j\mathbf{D},
\end{equation}
where $\mathbf{A}_j$ is an $M\times N$ binary matrix that we call the assignation matrix. It is a design parameter to select the binary sequences used by the $j$th \ac{UAV}, and also to determine in which branches they will be emitted. Thus, each row of the matrix $\mathbf{A}_j$ will have only a single entry with value '1'. We will briefly discuss the selection of the assignation matrix in section \ref{sec:discussion:txdesign}.
\par
We seek to design the dictionary matrix $\mathbf{D}$ for the system described in section \ref{sec:syst} to minimize the expected identification time for a fixed number of different optical signals, i.e. minimize the expected identification time given for a fixed number of rows of $\mathbf{D}$. We define the identification time of an optical signal as the time elapsed from its {\it birth time} (defined in section \ref{sec:syst:rx}) until the time when the classifier assigned to the signal successfully determines its identification number. The design of the assignation matrices $\{\mathbf{A}_j\}_{j=1}^J$ is beyond the scope of this article. Thus, they will be considered fixed with an arbitrary configuration.
\par
Regarding the encoder, we select the \ac{NRZ} coding as it maximizes the bit rate for the \ac{OOK} modulation (as long as synchronization problems are not considered) \cite{KwonIEEEPTL2010}. In section \ref{sec:detection:Manchester}, we briefly discuss utilization of the Manchester coding.

\section{Binary Sequences Construction and Combinatorial Analysis}
\label{sec:binarySeq}
The performance of the mutual vision-based identification system discussed in this article strongly depends on the set of binary sequences contained in the dictionary $\mathbf{D}$, as mentioned in section \ref{sec:syst:rx}. Let $\mathcal{X}^L$ be the set of all the binary sequences in the dictionary matrix $\mathbf{D}$ with dimensions $M\times L$. Let $\mathbf{b}_n$ denote the $n$th binary sequence in $\mathcal{X}^L$ and $\mathbf{b}_n[k]$ denote its $k$th bit, where $k=0,1,\cdots, L-1$. For simplicity, we disregard the effects of the clock signals mismatches in this section, but they will be studied theoretically in section \ref{sec:clock} and experimentally in section \ref{sec:experiments}.
\par
Before we proceed, we describe the requirements of the identification system receiver of Fig. \ref{fig:diagrams} (center). The image processing module in Fig. \ref{fig:diagrams} (center) must reliably detect the bright spots generated by the LEDs of the group's \acp{UAV}. It must also discriminate bright spots generated by the \acp{UAV} from the bright spots generated by random environmental lights. As the \acp{UAV} move and their relative positions change, the image processing module must track the motion of the blinking lights emitted by the LEDs of the \acp{UAV}. The classifier must determine the true identification number of the analyzed optical sequences as fast as possible. In addition, the identification system must support as many different optical signals as possible to enable its use for a large group of of \acp{UAV}.
\par
The requirements described above dictate the following requirements for the binary sequences in the set $\mathcal{X}^L$:
\par
1) To facilitate the detection and tracking of the bright spots generated by the LEDs, we ensure a minimum average power of the emitted optical signals. Since we are using the \ac{OOK} modulation with the \ac{NRZ} coding, the average power of the optical signal associated with the binary sequence $\mathbf{b}_n$ is proportional to the average power of the binary sequence. To ensure a minimum average power on all emitted optical sequences, we constrain all of the binary sequences to satisfy:
    \begin{equation}
    \label{eq1:7}
     \|\mathbf{b}_n\|_0\geq \bar{b}L,  
    \end{equation}
where $\bar{b}\in[0,1]$ is the desired normalized minimum average power, and $\|.\|_0$ is the $L_0$-norm.
\par    
2) Many bright spots on the camera frames that are not generated by \ac{UAV} LEDs are sunlight reflections. Some of these reflections are generated by static reflectors  and appear on many consecutive camera frames as a constant bright spots. We help to discriminate valid binary sequences from these reflections by limiting the maximum time that any LED can be continuously turned on. Thus, we limit to $N_1$ the number of circularly consecutive bits with value '1' for each binary sequence $\mathbf{b}_n$.
\par    
3) The image processing module must track the motion of all bright spots detected on the camera frame. We can implement the tracker using the Hough transform \cite{HartIEEESPM2009} as in \cite{UVDAR2} or {polynomial prediction as in \cite{Lakemann2025tracking} where the reader can find a detailed implementation and testing of the tracker}. But, regardless of the particular implementation, the general behaviour of the tracker is as follows. When the bright spot is detected on the camera frame, the tracker locks to the central pixel of the bright spot and starts tracking it. Since the \ac{UAV} LEDs are blinking, when the LED is turned off, the tracker must predict the central pixel of the bright spot, which should appear once the LED is turned on again. The longer the LED remains off, the larger the uncertainty of the central pixel location. If this uncertainty grows too large, the tracking will fail. To reduce the tracking failure, we limit the time that each LED can remain turned off by restricting $\mathbf{b}_n$ to have no more than $N_0$ {circularly consecutive} bits with value '0'.
\par
4) The emitted optical signals are periodic with a period of $L$ bits. The $L$ most recent bits received at the input of the classifier at time instant $k$, assuming no bit errors, are:
    \begin{equation}
    \label{eq1:8}
        \left[
        \begin{array}{cc}
        y[k]       \\
        y[k-1]\\
        \vdots\\
        y[k-L+2]\\
        y[k-L+1]
        \end{array}
        \right]=\left[
        \begin{array}{cc}
        \mathbf{b}_n[{\rm mod}(L-1+d,L)]       \\
        \mathbf{b}_n[{\rm mod}(L-2+d,L)]\\
        \vdots\\
        \mathbf{b}_n[{\rm mod}(d+1,L)]\\
        \mathbf{b}_n[{\rm mod}(d,L)]
        \end{array}
        \right],
    \end{equation}
    where $d$ is a random variable uniformly distributed within the discrete set $\{0,1,\dots,L-1$\}, representing the lack of time synchronization between the optical receiver and the optical transmitter. The classifier must identify $\mathbf{b}_n$, regardless of the random shift $d$ and without its knowledge. Thus, any two binary sequences $\mathbf{b}_n$ and $\mathbf{b}_m$ are considered equal if one is a circularly shifted version of the other, in which case we say that they are circularly equivalent. 
\par
5) When the \ac{SNR} is poor, the raw \ac{BER} is large, and can result in long identification times and constant identification failures. To alleviate this, we can add some robustness by increasing the circular Hamming distance of the set $\mathcal{X}^L$, defined as:
    \begin{equation}
    \label{eq1:11}
    D(\mathcal{\mathcal{X}^L})=\min_{\mathbf{b}_n,\mathbf{b}_m\in\mathcal{\mathcal{X}^L}}H_c(\mathbf{b}_n,\mathbf{b}_m).
    \end{equation}
where $H_c(\mathbf{b}_n,\mathbf{b}_m)\triangleq\min_d\left\|\mathbf{b}_n  \oplus c(\mathbf{b}_m,d)\right\|_0$ is the circular Hamming distance between the binary sequences $\mathbf{b}_n$ and $\mathbf{b}_m$; $c(\mathbf{b}_n,d)$ is the binary sequence $\mathbf{b}_n$ after being circularly shifted $d$ bits to the right; and $\oplus$ is the XOR logic operator. If $D(\mathcal{X}^L)=1$, then any single bit error can transform a valid binary sequence into another valid binary sequence. Thus, it is impossible to determine if the binary sequence was correctly decoded. If $D(\mathcal{X}^L)=2$, any single bit error will transform a valid binary sequence into an invalid binary sequence. Thus, it becomes possible to detect single bit errors, but it will not be possible to correct them. If $D(\mathcal{X}^L)=3$, then any single bit error will transform a valid binary sequence into an invalid binary sequence. However, the circular Hamming distance of this erroneous invalid binary sequence to the original binary sequence will be shorter than to any other valid binary sequence. Thus, it will be possible to detect and correct single bit errors. 

\subsection{Binary sequence set generation and analysis}
\label{sec:binarySeq:cardinality}
After establishing the sequence requirements, we construct $\mathcal{X}^L$ and study its cardinality. To do this, we use the algorithm \ref{algorithm1} with the following inputs: the set $\mathcal{S}^L$ of all the $2^L$ binary sequences of length $L$, the minimum value allowed for each sequence average $\bar{b}$ (see \eqref{eq1:7}), the maximum number $N_1$ ($N_0$) of {circularly consecutive bits} with value '1' ('0') for each sequence, and the circular Hamming distance $H_m$ for $\mathcal{X}^L$.
\par
\begin{algorithm}[h]
\caption{Sequences generation for \ac{NRZ} coding}\label{algorithm1}
\begin{algorithmic}[1]
\Procedure{$\mathcal{X}^L=f(\mathcal{S}^L, \bar{b}, N_1, N_0, H_m)$}{}
\State $\mathcal{A}^L$=PowerTest($\mathcal{S}^L,\bar{b}$)
\State $\mathcal{B}^L$=CircularityTest($\mathcal{A}^L$)
\State $\mathcal{C}^L$=OnesTest($\mathcal{B}^L,N_1$)
\State$\mathcal{D}^L$=ZerosTest($\mathcal{C}^L,N_0$)
\State $\mathcal{E}^L$=HammingTest($\mathcal{D}^L,H_m$)
\State \textbf{return} $\mathcal{E}^L$
\EndProcedure
\end{algorithmic}
\end{algorithm}
\par
We now discuss each step of algorithm \ref{algorithm1} and calculate the cardinality of the output set $\mathcal{X}^L$. To do this, we partition $\mathcal{S}^L$ into $L+1$ partitions, $\{\mathcal{S}^L_{\ell}\}_{{\ell}=0}^L$, where $\mathcal{S}^L_{\ell}$ is the partition containing all binary sequences $\mathbf{b}\in\mathcal{S}^L$ that satisfy $\|\mathbf{b}\|_0={\ell}$. The same partition is applied to each set. The cardinality of $\mathcal{S}^L_{\ell}$ is given by the binomial coefficient $L$ choose $\ell$:
\begin{equation}
\label{eq2:1}
    \mid\mathcal{S}_{\ell}^L\mid=\left(
\begin{array}{c}
      L  \\
      {\ell}
\end{array}
    \right).
\end{equation}
\subsubsection{Power test}
\label{sec:binarySeq:cardinality:power}
The power test sets the minimum power of the emitted optical signals to $\bar{b} P$. This is done by discarding the subsets $\mathcal{S}_{\ell}^L$ with ${\ell}< L \bar{b}$. Thus, the cardinality of $\mathcal{A}^L$ is:
\begin{equation}
\label{eq2:2}
\mid\mathcal{A}^L\mid=\sum_{{\ell}=\lceil L\bar{b}\rceil}^L\mid\mathcal{S}_{\ell}^L\mid.
\end{equation}
\subsubsection{Circularity test}
\label{sec:binarySeq:cardinality:circularity}
This test ensures that all binary sequences in $\mathcal{B}^L$ are circularly different. To do this, we extract a sequence $\mathbf{b}$ from $\mathcal{A}_{\ell}^L$, include it into $\mathcal{B}_{\ell}^L$, and eliminate from\footnote{All the circularly equivalent sequences to $\mathbf{b}\in\mathcal{A}_{\ell}^L$ have the same $L_0$ norm, and thus belong to the same partition.} $\mathcal{A}_{\ell}^L$ all of the sequences that are circularly equivalent to $\mathbf{b}$. We repeat this for each sequence in $\mathcal{A}_{\ell}^L$ until $\mid\mathcal{A}_{\ell}^L\mid=0$. Then, we repeat this process for all the remaining partitions of $\mathcal{A}^L$. Each binary sequence $\mathbf{b}\in\mathcal{A}_{\ell}^L$ has, at most, $L-1$ circularly equivalent sequences\footnote{It can have less if the sequence presents some symmetries.} in $\mathcal{A}_{\ell}^L$. Therefore, for $0<\ell<L$, we can approximate the cardinality of $\mathcal{B}_{\ell}^L$ as:
\begin{equation}
\label{eq2:3}
    \mid\mathcal{B}_{\ell}^L\mid\approx \mid\mathcal{A}_{\ell}^L\mid/L,
\end{equation}
while $\mid \mathcal{B}_0^L \mid=1$ and $\mid\mathcal{B}_L^L\mid=1$. 
\subsubsection{Ones and zeros tests}
\label{sec:binarySeq:cardinality:ones_zeros}
These tests ensure that each of the binary sequences has no more than $N_1$ {circularly consecutive bits} with value '1', and no more than $N_0$ {circularly consecutive} bits with value '0'. Let us start with the {\it Ones test}. All sequences within partitions $\{\mathcal{B}_{\ell}^L\}_{{\ell}\leq N_1}$ have no more than $N_1$ {circularly  consecutive} bits with value '1', since their $L_0$ norm is not larger than $N_1$ by definition. Consequently, $\mid\mathcal{C}_{\ell}^L\mid=\mid\mathcal{B}_{\ell}^L\mid$ for ${\ell}\leq N_1$. But, partitions $\{\mathcal{B}_{\ell}^L\}_{{\ell} > N_1}$ have sequences with more than $N_1$ {circularly consecutive} bits with value '1' and must be eliminated. To calculate $\{\mid\mathcal{C}_{\ell}^L\mid\}_{{\ell} \in (N_1,L-1]}$, we proceed as follows. Due to the circular equivalence, every single sequence $\mathbf{b}\in\mathcal{B}_{\ell}^L$ with $N_1<{\ell}<L$ having more than $N_1$ {circularly consecutive} bits with value '1' can be written, after some circular shifting, in the form of the following row vector:
\begin{equation}
\label{eq2:4}
\mathbf{b}=[1_{N_1+1},v_{L-N_1-2}(\ell-N_1-1),0],  \end{equation}
where $1_{x}$ is a binary row vector of length $x$ and $\|1_{x}\|_0=x$, $v_{x}(y)$ is any binary row vector of length $x$ with $\|v_{x}(y)\|_0=y$. The number of all sequences $\mathbf{b}\in\mathcal{B}_{\ell}^L$ in (\ref{eq2:4}) is determined by the number of different vectors $v_{L-N_1-2}(\ell-N_1-1)$, given by the binomial coefficient $L-N_1-2$ choose $\ell-N_1-1$:
\begin{equation}
\label{eq2:5}
    \Delta_{C}(\ell)=\left( 
    \begin{array}{c}
         L-N_1-2\\
          \ell-N_1-1
    \end{array}
    \right).
\end{equation}
The number of sequences $\mathbf{b}\in\mathcal{B}_{\ell}^L$ that violate the constraint of the maximum allowed number of {circularly  consecutive} bits with value '1' is approximately $\Delta_{C}(\ell)$. Thus, we have:
\begin{equation}
\label{eq2:6b}
    \mid\mathcal{C}_{\ell}^L\mid\approx \max\left(\mid\mathcal{B}_{\ell}^L\mid-\Delta_{C}({\ell}),0\right).
\end{equation}
\par
The {\it zeros test} is complementary to the {\it ones test} in algorithm \ref{algorithm1}. Thus, we use a similar procedure to estimate $\mid\mathcal{D}_{\ell}^L\mid$. 
\par
If $N_1\geq L-N_0$, then the {\it zeros-test} acts only on partitions that were not modified by the {\it ones-test}. Thus, using the same method used to derive (\ref{eq2:5})-(\ref{eq2:6b}), we obtain for the {\it zero-test}:    
\begin{eqnarray}
\label{eq2:7}
    \mid\mathcal{D}_{\ell}^L\mid&\approx& \max\left(\mid\mathcal{C}_{\ell}^L\mid-\Delta_{D}({\ell}),0\right),\\
\label{eq2:8}
    \Delta_{D}({\ell})&=&\left( 
    \begin{array}{c}
         L-N_0-2\\
          L-{\ell}-N_0-1
    \end{array}
    \right).
\end{eqnarray}
\par
If $N_1< L-N_0$, then we can divide the partitions into three groups: i) partitions affected by the {\it zeros-test} only, i.e., partitions that satisfy $\mathcal{C}_{\ell}^L=\mathcal{B}_{\ell}^L$ and $\mathcal{D}_{\ell}^L\neq\mathcal{C}_{\ell}^L$. These partitions are given by ${\ell}\in0,1,\cdots,N_1-1$; ii) partitions affected by both tests, i.e., partitions that satisfy $\mathcal{C}_{\ell}^L\neq\mathcal{B}_{\ell}^L$ and $\mathcal{D}_{\ell}^L\neq\mathcal{C}_{\ell}^L$. These partitions are given by ${\ell}\in N_1,N_1+1,\cdots,L-N_0$; and iii) partitions affected only by the {\it ones-test}, i.e., partitions that satisfy $\mathcal{C}_{\ell}^L\neq\mathcal{B}_{\ell}^L$ and $\mathcal{D}_{\ell}^L=\mathcal{C}_{\ell}^L$. These partitions are given by ${\ell}\in L-N_0+1,L-N_0+2,\cdots,L$. 
\par
The cardinality of $\{\mathcal{D}_{\ell}^L\}_{{\ell}\in[0,N_1)}$ is calculated using (\ref{eq2:7})-(\ref{eq2:8}). The cardinality of  $\{\mathcal{D}_{\ell}^L\}_{{\ell}\in(L-N_0,L]}$ remains the same as that of the partitions $\{\mathcal{C}_{\ell}^L\}_{{\ell}\in(L-N_0,L]}$. Regarding $\{\mathcal{D}_{\ell}^L\}_{{\ell}\in[N_1,L-N_0]}$, (\ref{eq2:7})-(\ref{eq2:8}) provide a poor cardinality estimation, as they disregard that some sequences that violate the {\it zeros-test} also violate the {\it ones-test} and thus were already discarded. After extensive numerical analysis, we derived the following heuristic approximation for the cardinality of partitions $\{\mathcal{D}_{\ell}^L\}_{{\ell}\in(L-N_0,L]}$:
\begin{eqnarray}
\label{eq2:10}
    \mid\mathcal{D}_{\ell}^L\mid&\approx& \max(\mid\mathcal{C}_{\ell}^L\mid- \Delta_{D}'({\ell}),0),\\
    \Delta_{D}'({\ell})&=&\max(\Delta_{D}({\ell})-\Delta_{C}({\ell}),0),
\end{eqnarray}
where $\Delta_{D}'({\ell})$ is based on the difference between the number of sequences eliminated by the {\it one-test} and those eliminated by the {\it zero-test}.
A good cardinality approximation for every partition $\mathcal{D}_{\ell}^L$ can be obtained by combining equations (\ref{eq2:2}), (\ref{eq2:3}), (\ref{eq2:5}), (\ref{eq2:6b}), (\ref{eq2:8}), (\ref{eq2:10}), and  (\ref{eq2:11}). We show the cardinality of $\mathcal{D}^L$ and its estimation, using the above-mentioned equations, in the fifth and sixth columns of Table \ref{tab:cardinalityHm3}, respectively.

\subsubsection{Hamming distance test}
\label{sec:binarySeq:cardinality:Hamming}
This test tries to maximize the cardinality of $\mathcal{X}^L$, while satisfying the circular Hamming distance $D(\mathcal{X}^L)=H_m$:
\begin{eqnarray}
\label{eq2:11}
\maximize_{\mathbf{f}} &\quad& \|\mathbf{f}\|_0
\end{eqnarray}
\begin{equation}
\label{eq2:12}
\begin{array}{l}
\mathrm{s.t.} \nonumber\\
H_m [\mathbf{f}]_k[\mathbf{f}]_j\leq H_c(\mathbf{b}_k,\mathbf{b}_j),\\ j,k=1,...,\mid\mathcal{D}^L\mid,\quad j\neq k    
\end{array}
\end{equation}
where $\mathbf{b}_k$ is the $k$th sequence in $\mathcal{D}^L$, and $\mathbf{f}$ is a binary vector of length $\mid\mathcal{D}^L\mid$ that indicates which sequences are included in $\mathcal{X}^L$. If $\mathbf{b}_k\in\mathcal{X}^L$, then $[\mathbf{f}]_k=1$. But, if $\mathbf{b}_k\notin\mathcal{X}^L$, then $[\mathbf{f}]_k=0$. (\ref{eq2:11}) describes a discrete combinatorial optimization problem. Verifying if a particular binary vector $\mathbf{f}^*$ constitutes an optimum solution requires the exploration of the full search space (composed of $2^{\mid\mathcal{D}^L\mid}$ elements). Thus, as $\mid\mathcal{D}^L\mid$ grows, the problem becomes more computationally expensive to solve. Consequently, less computationally demanding suboptimal solutions are of interest. 
\par
For $H_m=1$, we have that $\mathcal{X}^L=\mathcal{D}^L$. For $H_m=2$, we can derive a suboptimal solution by considering the following three properties: i) if $\mathbf{b}_k,\mathbf{b}_j\in\mathcal{D}^L_{\ell}$, then $H_c(\mathbf{b}_k,\mathbf{b}_j)\geq 2$ because $\|\mathbf{b}_k\|_0=\|\mathbf{b}_j\|_0$ and $H_c(\mathbf{b}_k,\mathbf{b}_j)> 0$. To transform any binary sequence $\mathbf{b}_k$ into any other sequence $\mathbf{b}_j$, at least two bit flips are needed: a 0 bit flip and a 1 bit flip. A single one bit flip on $\mathbf{b}_k$ would alter $\|\mathbf{b}_k\|_0$, and thus the resulting sequence would no longer belong to $\mathcal{D}^L_{\ell}$; ii) if $\mathbf{b}_j\in\mathcal{D}^L_{\ell}$ and $\mathbf{b}_k\in\mathcal{D}^L_{\ell\pm2}$, then $H_c(\mathbf{b}_j,\mathbf{b}_k)\geq 2$ due to $\|\mathbf{b}_j\|_0=\|\mathbf{b}_k\|_0\pm2$; and iii) if $\mathbf{b}_j\in\mathcal{D}^L_{\ell}$ and $\mathbf{b}_k\in\mathcal{D}^L_{\ell\pm1}$, then $H_c(\mathbf{b}_j,\mathbf{b}_k)\geq 1$ due to $\|\mathbf{b}_j\|_0=\|\mathbf{b}_k\|_0\pm1$. From these properties, we conclude that $D(\cap_{k}\mathcal{D}^L_{2k})\geq2$ and  $D(\cap_{k}\mathcal{D}^L_{2k+1})\geq2$, where $\cap$ is the intersection operator. Thus, a suboptimal solution for $H_m=2$ is:
\begin{equation}
\label{eq2:13}
    \mathcal{X}^L=\bigg\{
    \begin{array}{cc}
    \cap_{k}\mathcal{D}^L_{2k}     &  {\rm if}\, \mid\cap_{k}\mathcal{D}^L_{2k}\mid>\mid\cap_{k}\mathcal{D}^L_{2k+1}\mid\\
    \cap_{k}\mathcal{D}^L_{2k+1}     & {\rm otherwise}
    \end{array}
\end{equation}
and its cardinality is the sum of the cardinalities of the selected partitions, which we have already shown how to calculate. 
\par
Developing similar methods for $H_m>2$ is extremely complicated due to the growing complexity of the relations among the binary sequences and the partitions. A suboptimal solution can be obtained using methods based on random search. It is simple to implement, but difficult to analyze. 
\par
It is possible to analytically derive a coarse cardinality estimation for the optimum solution when $H_m=3$. To do this, we use a modified version of the definition of a sphere around a binary vector $\mathbf{c}$ as used in \cite{Pless1998}:
\begin{equation}
\label{eq2:15}
\mathcal{S}_r(\mathbf{c})\triangleq\{\mathbf{v}\in\mathcal{D}^L\, :\, H_c(\mathbf{c},\mathbf{v})\leq r\}
\end{equation}
where $\mathbf{c}$ is the center of the sphere of radius $r$. If $\mathbf{b}_j,\mathbf{b}_k\in\mathcal{X}^L$ with $D(\mathcal{X}^L)=3$, then $H_c(\mathbf{b}_j,\mathbf{b}_k)\geq 3$. The following properties also hold: i) the spheres of radius one of any two valid sequences do not overlap $\mathcal{S}_1(\mathbf{b}_j)\cap\mathcal{S}_1(\mathbf{b}_k)=\emptyset$; ii) the sphere of radius two of any valid sequences does not include any other valid sequence   $\mathcal{S}_2(\mathbf{b}_j)\cap\mathbf{b}_k=\emptyset$; iii) the spheres of radius two (or the larger) of any two valid sequences can overlap and so, in general, we have that $\mathcal{S}_n(\mathbf{b}_j)\cap\mathcal{S}_n(\mathbf{b}_k)\neq\emptyset$ for $n\geq 2$; and iv) regardless of the nature of $\mathcal{D}^L$ in (\ref{eq2:15}), the {\it volume} of the sphere of radius one of any sequence is bounded as follows: $\mid\mathcal{S}_1(\mathbf{b})\mid\leq L+1$. 
\par
From the properties described above, we can think, in an oversimplified manner, of the optimization problem \eqref{eq2:11}-\eqref{eq2:12} as the problem of forming as many spheres of radius one, defined by \eqref{eq2:15}, as possible while using the sequences in $\mathcal{D}^L$, where $\mathcal{X}^L$ is formed with the sequences that constitute the centers of all of the spheres. Following this reinterpretation of \eqref{eq2:11}-\eqref{eq2:12}, a coarse estimation for the cardinality of $\mathcal{X}^L$ when $D(\mathcal{X}^L)=3$ is the maximum number of spheres of radius one that can be formed with sequences in $\mathcal{D}^L$:  
\begin{eqnarray}
\label{eq2:19}
\mid\mathcal{X}^L\mid\approx \left\lceil{\mid\mathcal{D}^L\mid}/{(L+1)}\right\rceil.
\end{eqnarray}
In table \ref{tab:cardinalityHm3}, we plot, for a circular Hamming distance of 3, the cardinality of $\mathcal{X}^L$ in the seventh column and its estimate using (\ref{eq2:19}) in the eighth column. For $L\leq 11$, equation (\ref{eq2:19}) is accurate, but for larger values of $L$, it is poor.
\begin {table}[htp]
\caption {Cardinality Results} \label{tab:cardinalityHm3} 
\begin{center}
\addtolength{\tabcolsep}{-2pt}
\scalebox{1}{
\begin{tabular}{|c c c c | c c c c |} 
 \hline
 &  &  &  &  & &  &  \\
 $L$ & $\bar{b}$ & $N_1$ & $N_0$ & $\mid\mathcal{D}^L\mid$ & $\hat{\mid\mathcal{D}^L\mid}$& $\mid\mathcal{X}^L\mid$ & $\hat{\mid\mathcal{X}^L\mid}$ \\ [0.5ex] 
 \hline\hline
 8 & 0.1 & 6 & 4 & 29& 28 & 4 &4  \\ 
 \hline
 8 & 0.2 & 6 & 6 & 32 & 31& 5 &4 \\
 \hline
 8 & 0.5 & 3 & 7& 14 & 13 & 2 & 2 \\ 
 \hline
 10 & 0.3 & 7 & 3& 72 &70 & 8 &7 \\ 
 \hline
 10 & 0.4 & 3 & 6 & 56 & 54& 6 &6 \\
 \hline
 10 & 0.5 & 7 & 2 & 42& 41& 4 &4 \\ 
 \hline
 11 & 0.2 & 4 & 9 & 148& 148& 11 &${13}$ \\
 \hline
 11 & 0.2 & 4 & 3 & 97& 98& 9 &9  \\
 \hline
 11 & 0.2 & 6 & 8 & 172&172& 11 &${15}$ \\
\hline
 12 & 0.1 & 6 & 7 &326& 321& 20 &${26}$ \\
\hline
 12 & 0.4 & 3 & 7 & 159& 153& 13 &13 \\ 
 \hline
\end{tabular}
}
\end{center}
\end {table}


\section{Manchester coding}
\label{sec:detection:Manchester}

If we use Manchester coding instead of \ac{NRZ}, the encoder/modulator must operate twice as fast as the binary stream generator and the frequency division factor must be $d_f=2$ (see diagram in Fig. \ref{fig:diagrams}). In this case, each bit has a duration $T_b$ of two periods of the clock signal $c_j(t)$, i.e., $T_b=2T$. With Manchester coding, the LED always changes state in the middle of every bit. Consequently, regardless of the binary stream signal $s_{j,m}$, the average power of the emitted optical signal is $\mathbb{E}[v_{jm}(t)]=0.5P$. The LEDs will continuously be turned on for $2T$ at most, and also continuously turned off for $2T$ at most. Thus, the Manchester coding automatically satisfies some of the requirements listed in section \ref{sec:binarySeq}. Thus, if we use Manchester coding, we can drop some lines from Algorithm \ref{algorithm1} and use Algorithm \ref{algorithm2} instead.

\begin{algorithm}[ht]
\caption{Sequences generation for Manchester coding}\label{algorithm2}
\begin{algorithmic}[1]
\Procedure{$\mathcal{X}^L=f_M(\mathcal{S}^L, H_m)$}{}
\State $\mathcal{A}^L$=CircularityTest($\mathcal{S}^L$)
\State $\mathcal{B}^L$=HammingTest($\mathcal{A}^L,H_m$)
\State \textbf{return} $\mathcal{B}^L$
\EndProcedure
\end{algorithmic}
\end{algorithm}

Algorithm \ref{algorithm2} is less restrictive than Algorithm \ref{algorithm1}. It discards less sequences and, under the same conditions, generates a set $\mathcal{X}^L$ with higher cardinality. In other words, we require shorter sequence lengths $L$ to obtain the set $\mathcal{X}^L$ with a desired number of sequences. However, the bit duration $T_b$ when using the Manchester code is twice that of the \ac{NRZ} code. 
To compare both codes fairly, we note from Algorithm \ref{algorithm1} and Algorithm \ref{algorithm2} that $\mid f_M(\mathcal{S}^L, H_m)\mid$$=$$\mid f(\mathcal{S}^L, 0.5,2,2,H_m)\mid$. Using this equivalence, we generate various sets of binary sequences with Algorithms \ref{algorithm1} and \ref{algorithm2} for comparison. The result is shown in table \ref{tab:Manchester}. The left part of the table shows the results obtained by using Algorithm \ref{algorithm1} with the \ac{NRZ} coding, where $N_{\textrm{NRZ}}$ is the number of obtained sequences and $LT_b/T$ is the normalized sequence duration. On the right side of the table, we observe the same information for the sequences obtained using algorithm \ref{algorithm2} with the Manchester coding. We note that, for a given value of $N_{\textrm{NRZ}}$, the sequence duration is shorter. Thus, the \ac{NRZ} coding results in shorter sequence durations and consequently shorter identification times.

\begin {table}[ht]
\caption {\ac{NRZ} and Manchester Comparison} \label{tab:Manchester} 
\begin{center}
\addtolength{\tabcolsep}{-2pt}
\begin{tabular}{|c c c c | c c c c | } 
 \hline
 $N_{\textrm{NRZ}}$ & L & $H_m$ &$LT_b/T$&$N_{\textrm{Man}}$ & L  & $H_m$ &$LT_b/T$\\ [0.5ex] 
 \hline\hline
 6 & 8&1& 8 & 6&5 &1& 10 \\ 
 11 & 10&1& 10 & 12&6 &1& 12 \\
 24 & 12&1& 12 & 34&8 &1& 16 \\
 \hline
 7 & 14&3& 14 & 8&10 &3& 20 \\
 16 & 16&3& 16 & 18&12 &3& 24 \\
 28 & 18&3& 18 & 29&13 &3& 26 \\
 \hline
\end{tabular}
\end{center}
\end {table}

\par
Lastly, the use of optical signals of different frequencies as is done in \cite{UVDAR3,UVDAR4} is extremely inefficient. Using the \ac{FFT} we can demonstrate that with this strategy we can only produce $L/2$ different sequences of length $L$.


\section{Identification Time Analysis}
\label{sec:clock}
Next, we study the identification time of the sequences considering the clock signal impairments. We focus on the link from one LED of \ac{UAV}-1 to the camera of \ac{UAV}-0. For simplicity, we assume an errorless detection and perfect tracking of the bright spot generated by the LED.

\subsection{Ideal clock signals} 
\label{sec:detection:equalClock}
When all clock signals are stable (i.e., $n_{j,k}=0$ in \eqref{eq:clock}) and have the same true period (i.e., $T_j=T$), then (\ref{eq:clock}) becomes:
\begin{equation}
\label{eq:clock:1}
t_{j,k_j}=k_jT+t_{j,0},
\end{equation}
where $k_j$ is the local discrete-time index of the $j$th \ac{UAV}. From \eqref{eq:clock:1}, it is clear that the clock signals of the optical transmitter of the \ac{UAV}-1  and of the optical receiver of the \ac{UAV}-0 operate at exactly the same rate. Thus, the receiver always takes one sample per bit transmitted, and the only source of bit error detection comes from the noise discussed in section \ref{sec:syst:channel}.
\par
When $H_{m}=1$, the classifier must accumulate $L$ consecutive errorless bits to identify the binary sequence. In this case, the minimum detection time (normalized over  $T$) is $L$, which occurs when the first $L$ bits received are errorless. Given a bit error probability $p_b$, the following identification time $T_d$ probabilities hold:
\begin{eqnarray}
\label{eq:clock:7}
    Pr(T_d<L)&=&0,\\
\label{eq:clock:8}
    Pr(T_d=L)&=&(1-p_b)^{L}.
\end{eqnarray}
For $Pr(T_d=L+m)$ with $m=1,\cdots,L$, the most recent $L$ bits received must be errorless and the $m-L$ bit (counted backwards) must be erroneous. Therefore, we have:
\begin{equation}
\label{eq:clock:9}
    Pr(T_d=L+m)=p_b(1-p_b)^{L}.
\end{equation}
For $m\in1,2,\cdots,L-1]$, we have
\begin{eqnarray}
\label{eq:clock:10}
    Pr(T_d=2L+m)&=&1-p_b(1-p_b)^{2L}\nonumber\\
    &-&(m-1)p_b^2(1-p_b)^{2L},    \nonumber\\
\end{eqnarray}
and for $m\geq L$, we have
\begin{eqnarray}
\label{eq:clock:10b}
    Pr(T_d=2L+m)&=&\left(1-\sum_{k=0}^{m-1}Pr(T_d=L+k)\right)\nonumber\\
    &\times& p_b(1-p_b)^{L}.    
\end{eqnarray}
From \eqref{eq:clock:9}-\eqref{eq:clock:10b}, the expected identification time is:
\begin{eqnarray}
\label{eq:clock:10c}
    \mathbb{E}[T_d]
    &=&\left[\frac{3p_b}{2}L^2+\left(1+\frac{p_b}{2}\right)L\right](1-p_b)^L\nonumber\\
    &+&\left[\sum_{n=2L+1}^{\infty}n\left(1-\sum_{k=L}^{n-L-1}Pr(T_d=k)\right)\right]\nonumber\\
    &\times&(1-p_b)^L
\end{eqnarray}
From \eqref{eq:clock:10c}, we generally observe that the expected identification time is a strictly increasing nonlinear function of the sequence length $L$. This analysis also holds for $H_{m}=2$, but for $H_{m}=3$, the analysis is more complex.
\par

\subsection{Stable clock signals with uncertain oscillation frequency}
\label{sec:detection:diffClock}
A more realistic case is when all clock signals are stable, but we consider the uncertainty due to the fabrication process in the clock period (i.e., $\sigma^2_T\neq0$). Then, \eqref{eq:clock} becomes:
\begin{equation}
\label{eq:clock2:1}
t_{j,k_j}=k_jT_j+t_{j,0}.
\end{equation}
After doing some algebra in \eqref{eq:clock2:1} and \eqref{eq:snr:2b}, $k_t$ becomes: 
\begin{equation}
\label{eq:clock2:5}
k_t=\left\lfloor\frac{\Delta+k_0T_0}{T_1}\right\rfloor
-\left\lceil\frac{2\Delta_{t}}{T_1+T_0+\mathrm{sign}(\Delta_{t})(T_1-T_0)}\right\rceil,
\end{equation}
where $\Delta_{t}=t_{0,0}-t_{1,0}$ and
\begin{equation}
\label{eq:clock2:4}
\Delta=\Delta_{t}
-\left\lfloor\frac{2(\Delta_{t})}{T_1+T_0+\mathrm{sign}(\Delta_{t})(T_1-T_0)}\right\rfloor.
\end{equation}
The index $k_t$, defined in \eqref{eq:snr:2b}, is the value of the local discrete time index $k_1$ at the transmitter when the local discrete time index at the receiver is $k_0$. In other words, at local discrete time $k_0$, the receiver samples the $k_t$th bit emitted by the receiver.  
\par
Let us define the following random variable: 
\begin{equation}
\label{eq:clock2:6}
\delta_{0,1}\triangleq {T_0}/{T_1}-1.
\end{equation}
As mentioned in section \ref{sec:syst:clock}, $T_{j_0}$ and $T_{j_1}$ are statistically independent. We also assume that\footnote{This is a realistic assumption for oscillators of reasonable quality.} $\sigma^2_T/T^2\ll 1$, and that the skewedness of $T_j$ is zero, i.e., its probability distribution is symmetric w.r.t. its mean. Using the Taylor series approximations in \eqref{eq:clock2:6}, we demonstrate that $\mathbb{E}[\delta_{0,1}]\approx0$ and that:
\begin{equation}
\label{eq:clock2:6b}
\mathrm{var}[\delta_{0,1}]=\sigma^2_{\delta}\approx 2\sigma^2_T/T^2.    
\end{equation}
\par
The r.h.s. of (\ref{eq:clock2:5}) is composed of a time-variant term and of a constant term. We rewrite the first term using (\ref{eq:clock2:6}):
\begin{equation}
\label{eq:clock2:7}
\left\lfloor{(\Delta+k_0T_0)}/{T_1}\right\rfloor
=k_0+\left\lfloor{\Delta}/{T_1}+k_0\delta_{0,1}\right\rfloor.
\end{equation}
We observe in (\ref{eq:clock2:7}) that the time variant term of
$k_t$ in (\ref{eq:clock2:5}) is composed of a linear term $k_0$ and a nonlinear function of $\delta_{0,1}$ and $k_0$. When the receiver's clock is slower than the transmitter's clock, we have $\delta_{0,1}>0$, and the nonlinear function in (\ref{eq:clock2:7}) increases one unit approximately every $\lfloor1/\delta_{0,1}\rfloor$ sampling instants. However, when the receiver's clock is faster than the transmitter's clock, we have $\delta_{0,1}<0$, and the nonlinear function in (\ref{eq:clock2:7}) decreases one unit approximately every $\lfloor1/\mid\delta_{0,1}\mid\rfloor$ sampling instants. Consequently, every $\approx\lfloor1/\mid\delta_{0,1}\mid\rfloor$ sampling instants, the receiver will miss a bit (if $\delta_{0,1}<0$) or duplicate a bit (if $\delta_{0,1}>0$). If the length of the emitted binary sequence, $L$, is larger than $\lfloor1/\mid\delta_{0,1}\mid\rfloor$, then the sequence will always be received with a missing or a duplicated bit. 
Thus, to ensure correct reception,  
the binary sequence length must satisfy $L<\lfloor1/\mid\delta_{0,1}\mid\rfloor$. Since $\delta_{01}$ is actually a random variable, the probability that the transmission of a binary sequence with length $L$ fails every time is $\mathrm{Pr}\left(L\geq \mid\delta_{01}\mid^{-1}\right)$. From \eqref{eq:clock2:6b} and the Chebyshev's inequality, we bound this probability as:
\begin{equation}
\label{eq:clock2:9}
    \mathrm{Pr}\left(L\geq \mid\delta_{01}\mid^{-1}\right)\leq \sigma^2_{\delta}L^2={2\sigma^2_{T}L^2}/{T^2},
\end{equation}
The probability that transmission through this link will work correctly is the complement to (\ref{eq:clock2:9}).
The resulting inequality is the key to determining the maximum sequence length $L$ for a group of $J$ \acp{UAV}. From \eqref{eq:clock2:9}, the probability that all possible $J(J-1)$ links can operate correctly is:
\begin{eqnarray}
\label{eq:clock2:11}
p_{g}\left(J\right)&=&    \mathrm{Pr}\left(\delta_J< L^{-1}\right)
\end{eqnarray}
The random variable $\delta_J$ is the maximum value of the $J(J-1)$ identically distributed random variables $\{\mid\delta_{j,k}\mid\}_{j\neq k}$. However, that set of $J(J-1)$ variables is generated by only $J$ independent random variables ($\{T_j\}_{j=1}^J$). Thus, the random variables $\{\mid\delta_{j,k}\mid\}_{j\neq k}$ are not statistically independent. But, if we neglect this, then we can make the following approximation:
\begin{equation}
\label{eq:clock2:13}
    p_{g}\left(J\right)\approx\left(\mathrm{Pr}\left(\mid\delta_{jk}\mid< L^{-1}\right)\right)^{J(J-1)},
\end{equation}
Further, using \eqref{eq:clock2:9}, we obtain:
\begin{eqnarray}
\label{eq:clock2:14}
    p_{g}\left(J\right)\geq\left(1-{2\sigma^2_{T}L^2}/{T^2}\right)^{J(J-1)}.
\end{eqnarray}
The probability $p_{g}\left(J\right)$ that all optical links operate correctly in the group of $J$ \acp{UAV} is lower bounded according to \eqref{eq:clock2:14}. Consequently, if we want all the optical links in a group of $J$ \acp{UAV} to work correctly with a probability that is lower bounded by $p_{g}\left(J\right)$, then the sequence length must satisfy:
\begin{equation}
\label{eq:clock2:15}
L_{max}\triangleq \frac{T}{\sigma_{T}}\sqrt{\frac{1}{2}\left(1-\exp\left(\frac{\ln{(p_{g}\left(J\right))}}{J(J-1)}\right)\right)}\geq L.
\end{equation}
The maximum sequence length $L_{max}$ is proportional to the nominal clock period $T$ and inversely proportional to the standard deviation of the clock period $\sigma_{T}$. $L_{max}$ is a decreasing function of the number of \acp{UAV} within the group. \par

{
\subsection{Unstable clock frequency}
\label{sec:detection:unstableClock}
We consider now the case when all clock signals have the same nominal frequency, but there is frequency instability (i.e., $n_{j,k}\neq0$).%

In this case, the time between falling edges of the clock signal becomes time-variant, and so the time between errors due to missing/duplicated bits becomes also time-variant. This means that sometimes the receiver's clock is faster than the transmitter's one and hence bits get duplicated. While some other times the receiver's clock is slower than the transmitter's clock and hence bits are missed. Our experiments showed that the p.d.f. of $n_{j,k}$ is symmetric, see Fig. \ref{fig:camclock1}, so missing and duplicated bits are equiprobable. Furthermore, simulations showed that, in general, missing and duplicated bits have the same effect on the \ac{BER}.

Now, clocks of reasonable quality have low frequency instability (e.g., in our experiments the time between frames was contained within $\pm 0.3\%$ of the mean). Thus, in practice the temporal variation of the time between errors is small, as observed in \ref{fig:staticExp1} where it varies mostly 1 or 2 samples around the mean. Hence, a practical design strategy is to initially assume $n_{j,k}=0$ (i.e., no frequency instability), then calculate the time between errors as done in section  \ref{sec:detection:diffClock}, then subtract to the time between errors a few samples to create a conservative estimate and then use it design  purposes.

\subsection{Exposure time effect}
For simplicity, up to now in section \ref{sec:clock} we assumed the bit sampling to be instantaneous. But its duration corresponds to the exposure time $\tau_e$. To conclude the study presented in this section we briefly discuss the effect of the non-zero $\tau_e$.

As pointed in (\ref{eq:snr:2}), when the sampling starts close to a the transition of two bits then the sample results in a mixture of both consecutive bits. The consequence is that, in the scenario of section \ref{sec:detection:diffClock} when we are close to the event of a missing/duplicated bit then there are a number of samples which are actually mixtures of consecutive bits, this number is given by
\begin{equation}
\label{eq:E_cont}
    E_{cont}=\left\lceil\frac{\tau_e}{\mid T_0-T_1 \mid}-1\right\rceil.
\end{equation}
In general, these samples are more prone to be misdetected due to their mixed nature. When there is frequency instability this effect still occurs but $E_{cont}$ becomes time variable and it is not given anymore by (\ref{eq:E_cont}).}
\section{Simulations and Experiments}
\label{sec:experiments}

\subsection{Hamming distance effect on the identification time}
\label{sec:experiments:hamming}
We consider a group of $J=11$ \acp{UAV} with two different blinking sequences assigned to each \ac{UAV}, i.e. $\mathrm{dim}(\mathbf{S}_j)=2$. Thus, we need to construct a dictionary $\mathbf{D}$ with at least 22 different binary sequences. In addition, we set $\bar{b}=0.4$, $N_1=7$, and $N_0=7$. Next, we consider two different cases. In case $A$, we construct a set with $H_m=1$, and sequence length $L=8$ (the minimum length that satisfies the desired number of sequences) which contains a total of 22 different sequences. In case $B$, we construct a second set with $H_m=3$, and sequence length $L=13$ (the minimum length that satisfies the desired number of sequences) which contains a total of 22 different sequences. 
\par
For both cases, we perform simulations to calculate the classification probability error $p_{ce}$ and the expected identification time $T_d$ for different bit probability errors $p_b$. We perform these simulations assuming $\sigma_T=0$, i.e., perfect clock signals. The results are presented at the top of Table \ref{tab:Results1}. For $p_b\leq0.2$, the identification time $T_d$ for case $B$ is longer, but its classification error probability is lower. In the absence of a clock signal mismatch, the benefits of the reduced classification error probability provided by the robustness obtained by increasing the circular Hamming distance weaken as the bit probability error increases. This is because sets of sequences with a larger circular Hamming distance must have larger lengths to maintain their cardinality, thus presenting more errors. 
\begin {table}[ht]
\caption {Simulation Results}
\label{tab:Results1} 
\begin{center}
\addtolength{\tabcolsep}{-1pt}
\begin{tabular}{|c |c |c| c |c | } 
 \hline
    $ (\delta=0), p_b$  & $2\cdot 10^{-1}$ &  $10^{-1}$ &$10^{-2}$&$10^{-3}$  \\ \hline   
    $\mathbb{E}[T_d]$ (case $A$) 
 &  21.404 &   12.751 & 8.369    & 8.031 \\ 
    $\mathbb{E}[T_d]$ (case $B$) 
 &  24.927&  15.598   & 13.025    &  13.001 \\ \hline
    $p_{ce}$ (case $A$) 
 & 0.789 &  0.538  & 0.073    &  0.0070 \\    
    $p_{ce}$ (case $B$)
 & 0.687 & 0.322 & 0.006   &   0.0001\\ [0.5ex] 
 \hline
 \hline
    $(\mid\delta\mid=0.01),p_b$  & $2\cdot 10^{-1}$ &   $10^{-1}$&    $10^{-2}$&$10^{-3}$  \\ \hline   
    $\mathbb{E}[T_d]$ (case $A$)
 & 21.735&  13.036 & 8.542    & 8.205 \\ 
    $\mathbb{E}[T_d]$ (case $B$)
 & 25.025 &  16.134  &13.267   &  13.205 \\ \hline
    $p_{ce}$ (case $A$) 
 & 0.795 &  0.559  &0.120    & 0.059 \\    
    $p_{ce}$ (case $B$)
 & 0.704& 0.368 & 0.064  &   0.057 \\ [0.5ex] 
 \hline
\end{tabular}
\end{center}
\end {table}

\subsection{Clock Period Uncertainty effect on the maximum identification capacity}
\label{sec:experiments:clockdiff}

Running the same simulations for the same set of binary sequences as was done in section \ref{sec:experiments:hamming}, we now consider a mismatch between the transmitter and receiver clocks of $\delta=0.01$ (i.e., one bit missed every 100 bits transmitted approximately) and present the results at the bottom of Table \ref{tab:Results1}. 
\par
We observe a slight increase in the identification time, but the classification error probability has a more interesting behaviour: when the bit error probability is high, the classification error probability for case $A$ gets close to that of case $B$, just slightly higher. When the bit error probability is medium, the classification error probability is significantly lower for  case $B$. When the bit error probability is low, the classification error probability for both cases reaches a common lower limit, where the errors due to the clocks mismatch dominate the errors due to the individual bit decoding errors. Thus, increasing the \ac{SNR} (which implies a further decrease of $p_b$) will not contribute to further reducing the classification error probability. When $p_b$ is low, the classification error probability for case $A$ is slightly lower than that for case $B$, because longer binary sequences are more affected by the clocks mismatch. 

    \begin{figure}[htp]
    \centering
      \includegraphics[width=1.0\linewidth]{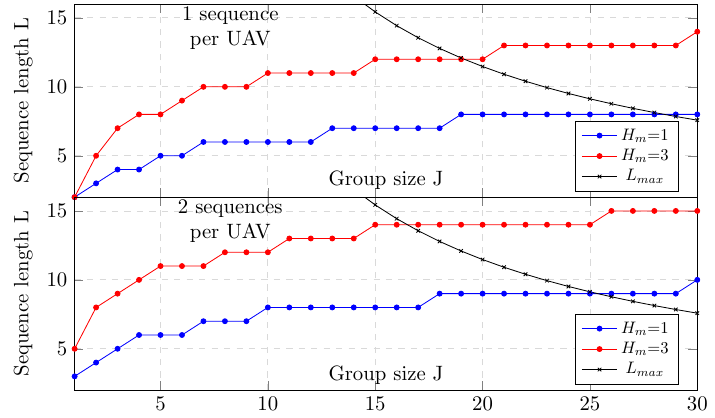}
        \caption{In black, $L_{max}$ for $p_g(J)$=0.999 and $T/\sigma_T=10^{4}$.In blue and red, the minimum sequence length $L$ of a set with circular Hamming distance 1 and Hamming distance 3, respectively, that satisfies the cardinality constraints for a group of $J$ \acp{UAV}.}
        \label{fig:limitAnalysis}%
    \end{figure}
    
As mentioned in section \ref{sec:detection:diffClock}, the variance $\sigma_T^2$ of the clock used in the \acp{UAV} determines the capacity of the identification system. To illustrate this, we analyse groups composed of $J$ \acp{UAV} equipped with reasonably accurate clocks ($T/\sigma_T=10^{4}$). In Fig. \ref{fig:limitAnalysis}, we plot in black the maximum sequence length $L_{max}$ that can be used by the identification system, so that we have a probability $p_{g}\left(J\right)$ that  all optical links can operate correctly (i.e., that any \ac{UAV} can identify any other \ac{UAV}). In blue, we plot the minimum sequence length $L$ of a set with circular Hamming distance 1, required to assign one sequence per \ac{UAV} (top) and two sequences per \ac{UAV} (bottom); and in red, we plot the minimum sequence length $L$ of a set with $H_m=3$. 
\par
In the top image, we observe that for $L\geq 29$, both the blue and red curves are above $L_{max}$. This means that given the clocks used by the \acp{UAV}, it is not possible to assign one distinct sequence per \ac{UAV} and ensure that any \ac{UAV} in the group can identify any other \ac{UAV} in the group with a probability of $p_{g}\left(J\right)$ or higher. In the bottom image where we assign two different sequences per \ac{UAV}, this occurs for $L\geq 26$. On the left, for $20 \leq J\leq 28$ only, sets of sequences with $H_m=1$ can ensure the proper operation of all optical links. For $J\leq 19$, we can use either sets of sequences with $H_m=1$ or $H_m=3$. 

\subsection{Camera interframe duration analysis}
\label{sec:experiments:clockSignalAnalysis}
We recorded the camera interframe duration of the \ac{UAV} shown in Fig. \ref{fig:3_uav_platforms} during 400 seconds.
The timings were recorded as those reported by the camera driver\footnote{\texttt{\url{https://github.com/ctu-mrs/bluefox2}}} used on the platforms.
The nominal camera frame rate was set to 60 frames/s.
The p.d.f. derived from the measurements histogram was recorded and compared to the  Laplace distribution with its parameters estimated using maximum log-likelihood. We observed an excellent match between both distributions, and thus we conclude that the camera interframe duration follows a Laplace probability distribution, {see Fig. \ref{fig:camclock1}. In Fig. \ref{fig:camclock2}, we observe the power spectrum of the temporal variations of the camera interframe duration measured in the experiments.We observe that the power spectrum of the temporal variations of the camera interframe duration is not flat and so the term $n_{j,k}$ in (\ref{eq:clock}) cannot be modelled as a white noise process. 

    \begin{figure}[htp]
    \centering
      \includegraphics[width=\linewidth]{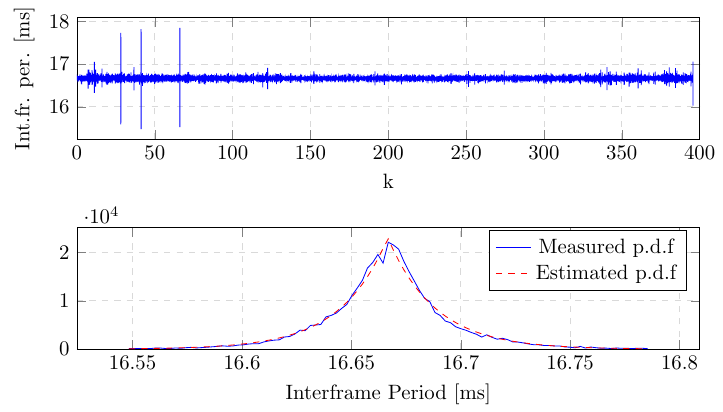}
        \caption{Interframe camera duration recorded (top). The p.d.f. of the interframe camera duration (bottom).  }
        \label{fig:camclock1}%
    \end{figure}
    
    \begin{figure}[htp]
    \centering
      \includegraphics[height=3cm,width=\linewidth]{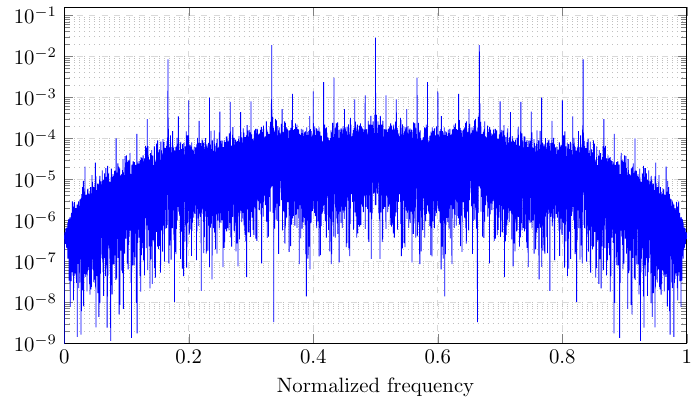}
        \caption{Power spectrum of the interframe camera period temporal variations.}
        \label{fig:camclock2}%
    \end{figure}}

To continue investigating the effect of the clock signal impairments, we perform the following indoor experiment.
We place two \acp{UAV} (\ac{UAV}-0 and \ac{UAV}-1) of the same model as before \SI{5}{\meter} apart on the floor of the laboratory. The \ac{UAV}-0 camera operates with a nominal frame rate of 60 frames/s, and points to \ac{UAV}-1. \ac{UAV}-1 points one arm towards the \ac{UAV}-0 camera, while two adjacent arms remain oriented parallel to the \ac{UAV}-0 camera image plane. We name these LEDs, from left to right on their camera image, as D0, D1, and D2.

The bit-rate of the blinking LEDs was \SI{60.241}{\hertz} due to hardware limitations of the microcontroller architecture used in the LED driver.
This creates a difference in the period of less than \SI{66.7}{\micro\second} compared to the nominal camera period of \SI{16.167}{\milli\second}, with the exposure time of the camera set to \SI{500}{\micro\second}.
The precision of the crystal clock of the LED driver incurs an additional error several orders of magnitude lower than the other sources of error.
Each LED emits an optical signal generated with a different binary sequence. We generate a set of sequences $f(\mathcal{S}^7, 0.5, 3, 3, 1)$ using Algorithm \ref{algorithm1}. We then feed the binary stream generators (see Fig. \ref{fig:diagrams} (right)) associated with LEDs D0, D1, and D2 the sequences 0010111, 0011011, and 0011101. Then, we record the \ac{UAV}-0 camera footage for \SI{180}{\second}.
\par
Due to the short distance between the \acp{UAV}, the pixels corresponding to the light emitted by the LEDs get almost saturated when the LEDs are turned on. Thus, the \ac{SNR} at the receiver is high and the effects of the clock signal impairments and mismatch dominate over the effects of the noise. 
\begin{figure}[htp]
    \centering
    \includegraphics[width=1\linewidth]{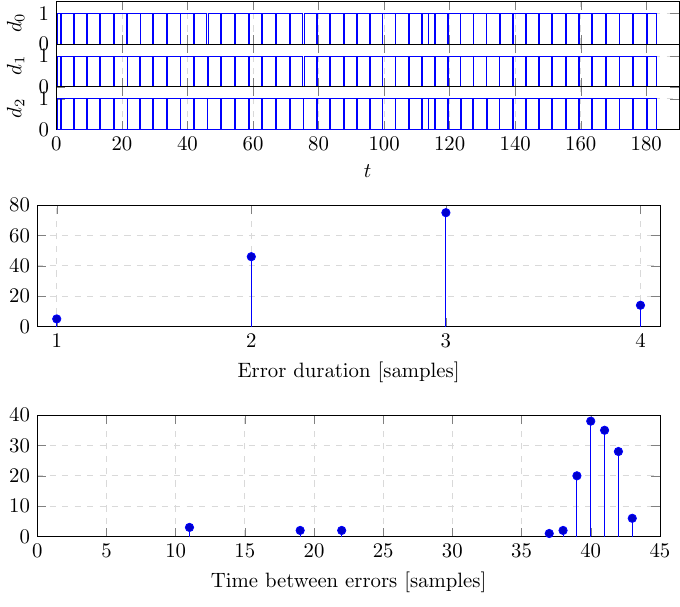}
    \caption{The time progression and error analysis classification success signals for $f(\mathcal{S}^7, 0.5, 3, 3, 1)$ at \SI{5}{\meter}.}
    \label{fig:staticExp1}
    \label{fig:staticExp2}
\end{figure}\par
In Fig. \ref{fig:staticExp1}, we plot, for each LED, a binary signal that takes the value 1 when the binary sequence embedded in the optical signal is correctly classified and 0 otherwise. In these plots, we note the presence of quasi periodic errors, which are generated by the mismatch between the transmitter and the receiver clock signals, as described in section \ref{sec:clock}. 

To analyze these errors in more detail, we plot the histograms of their duration and of the time between the errors measured from start-to-start in Fig. \ref{fig:staticExp2}. From these histograms, we note two different types of errors: the first type of error is caused by noise at the receiver, and it has a duration of one single sample. The time between them does not follow any specific pattern; the second type of error is due to missing/duplicated bits caused by clock mismatches. In this experiment, they can last between two and four samples. The time between them is a random variable with a mean of around 40.6 samples and a variance of around 1.49 samples. The histograms in Fig. \ref{fig:staticExp2} show the following: i) if the optical signals are generated with sequences with a length of 39 bits or larger, they will always be received with errors; ii) the clock signal mismatch is time-variant, which is why the time between errors caused by missing/duplicated bits varies mostly between 39-42 samples; and iii) at high \ac{SNR}, the main factor that limits the performance of the optical identification system is the mismatch between the transmitter and receiver clocks.

\subsection{Dynamic outdoor testing}

As proof of concept, we deployed a group of three \acp{UAV} (\ac{UAV}-0, \ac{UAV}-1, and \ac{UAV}-2) equipped with the \ac{UVDAR} system outdoors.
We used binary sequences taken from the set generated with $f(\mathcal{S}^{14}, 3, 3, 0.5, 1)$ and assigned one sequence to each arm of each \ac{UAV}, where each \ac{UAV} emitted four unique optical signals through its LEDs. Fig. \ref{fig:uvdar_view_experiment} shows a camera snapshot with correctly classified markers. 
The \acp{UAV} flew autonomously according to a formation enforcement technique developed in our laboratory.
The ability to distinguish between the individual signals allowed the \acp{UAV} to identify each other and estimate their relative orientations. This flight allowed us to test the visual identification system in a more challenging and realistic scenario.

We recorded the content of the \ac{UAV}-0 camera for \SI{235}{\second} with a nominal frame rate of 60 frames/s.
The clock signal of the optical transmitters of all the \acp{UAV} operate with a nominal frequency of \SI{60}{\hertz}.
The trajectories of the \acp{UAV} and the execution of the identification process are shown in a video of the experiment at the following link\protect\footnotemark.

    \par
We evaluated the classification success signal for each individual LED of the \ac{UAV}-1 and of the \ac{UAV}-2  by the left camera of \ac{UAV}-0.
This signal takes on a value of one when the classification is successful and takes a zero value when there is an error in classification, or when the LED leaves the \ac{UAV}-0 camera's \ac{FoV}.
From the video of the experiment, we observe that usually only two LEDs per \ac{UAV} are captured by the \ac{UAV}-0 camera, although sometimes three LEDs can be captured simultaneously.
We also note that, \ac{UAV}-2 leaves the \ac{UAV}-0 camera's \ac{FoV} and is lost for some moments. The detection success is significantly more erratic than in the prior testing with the static transmitter and receiver, since the motion of the \acp{UAV} affects the optical signal retrieval.
Additionally, the distances between the transmitters and receivers were at times greater, and the contrast of the active LEDs in the image was slightly lower due to sunlight.
Despite this, the classification success was sufficient for the formation enforcement system to perform its function, which testifies to the practical applicability of the proposed identification system in real-world conditions.

\begin{figure}[htp]
    \centering
    \includegraphics[width=1.0\linewidth]{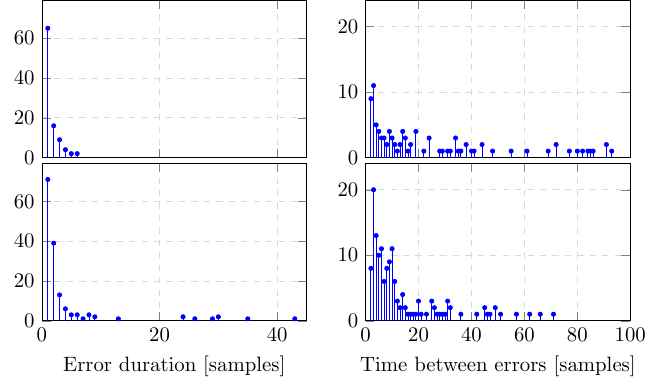}
    \caption{Error duration and time between errors histograms for \ac{UAV}-1 (top) and \ac{UAV}-2 (bottom).}
    \label{fig:trajUAVhistDuration}
    \label{fig:trajUAVhistDelay}
\end{figure}\par

\par
For both signals mentioned above, the histograms of the duration of the errors observed are plotted in Fig. \ref{fig:trajUAVhistDuration}. In Fig. \ref{fig:trajUAVhistDelay}, we plot the histograms of the time between the errors. The first thing to note in Fig. \ref{fig:trajUAVhistDuration} is that most of the errors for both \acp{UAV} last only a single sample. For \ac{UAV}-1, single sample errors constitute 66.33$\%$ of the total errors. For \ac{UAV}-2, single sample errors constitute 47.02$\%$ of the total errors. From Fig. \ref{fig:trajUAVhistDelay}, it can be observed that errors occur more often; this is due to the larger distances (lower \ac{SNR}) and the additional effect of the blurring and tracking errors. Despite the challenging conditions, the probability of correctly detecting \ac{UAV}-1 is 0.9311, and 0.6327 for detecting \ac{UAV}-2. Further, the errors appear in short bursts as long as the \ac{LoS} is present. This demonstrates that our identification system performs well in real scenarios.

\footnotetext{\footnotesize{\texttt{http://mrs.felk.cvut.cz/uvdar-identification-sequences}}}

{
\section{Discussion}
\label{sec:discussion}
Before concluding, in this section we discuss some important aspects of the optical transceiver presented in Fig. \ref{fig:diagrams}.
\subsection{Optical Transceiver design}
\label{sec:discussion:txdesign}
In this paper we focus on the optical transmitter and receiver of Fig. \ref{fig:diagrams}. The receiver deals with individual markers, processed in parallel but independently, and so the specific choice of the assignation matrices $\{\mathbf{A}_j\}_{j=1}^J$ introduced in section \ref{sec:Problem} is irrelevant for its performance. Nevertheless, the criteria for selecting $\{\mathbf{A}_j\}_{j=1}^J$ can come from the upper layers. For instance, the upper layers might select $\{\mathbf{A}_j\}_{j=1}^J$ so that the circular Hamming distance between the sets $\mathbf{S}_j$ and $\mathbf{S}_k$ is maximized for all $k\neq j$, see (\ref{eq:00S}). This could be done in order to reduce the probability of mistaking one \ac{UAV} with another.  

Now, regarding the optical transmitter, as the number of branches $M$ increases the upper layers can extract more information, but this can also reduce the performance of the optical receiver due to shorter distances between the markers on the \acp{UAV}. This is because, if the markers on the \ac{UAV} emit different sequences, they will create mutual interference which will make it harder for the receiver to identify correctly each sequence. But, note that this degradation does not occur if all markers on the \ac{UAV} emit the same optical signal. In addition, increasing the number of branches $M$ in the optical transmitter requires more hardware and more processing at the receiver since more markers have to be identified per \ac{UAV}.

\subsection{Camera}
\label{sec:discussion:camera}
Let us start the discussion about camera by explaining the use of \ac{UV} light instead of near \ac{IR} light. 
The cameras used are off-the-shelf greyscale cameras intended for visible light, but they are also sensitive to nearby \ac{UV} and \ac{IR} wavelengths, and the sensitivity of the sensor decreases in both directions. In the solar spectrum, the intensity of \ac{UV} decays fast with decreasing wavelength near the visible range, while the \ac{IR} decays with increasing wavelength, but significantly more gradually. 
Additionally, objects contained in a typical outdoor environment tend to be less reflective to \ac{UV} than to both visible light and to \ac{IR}.
Because of these effects, it was possible to use near-\ac{UV} band-pass filters on our cameras to look into a radiation range with very dark background, while retaining a good sensitivity of the camera to our artificial markers mounted on the \acp{UAV}. More details on the reasons to use \ac{UV} light in this system can be found in our previous work \cite{UVDAR1}.
If we use bandpass filters and emitters of \ac{IR} instead, we would have to deal with a stronger trade-off between background light contamination and the distance from visible light we focus on, with which the camera sensitivity decreases.

Now, the camera setup we used consists of three cameras laid out as in Fig. \ref{fig:uav_platform}.
Each of these has a fisheye lens, which in combination with the camera chip we use covers approximately 180 degrees horizontally and 120 degrees vertically, with slight visibility behind the camera in the corners.
Two of these cameras are laid out 70 degrees rotated horizontally from the front of the \ac{UAV}, such that the longer side of the camera chips are horizontal.
The last one is pointed backwards, and with the longer side of the chip oriented vertically. 
This way, the horizontal plane centered on the UAV is fully covered, with overlaps.
This configuration presents some here are some roughly triangular blind spots located above and below the front of the \ac{UAV}. In addition, the \ac{UAV}'s own body create some small obstructions to the camera which result in minor blind spots.
One way this can be addressed is by using cameras with smaller chips, that will thus cover larger portion of the lenses projection circle, or by using four cameras with a different layout.

To finalize this subsection we have to mention that the effective range of our optical identification system with the specific setup used (i.e., camera chips, lenses, chosen exposure rate, LED markers, etc.), is approximately 15 meters.
This range is the result of various considerations, including the distance at which we can reliably resolve a individual \ac{UV} LEDs with our cameras and the power emitted by the LEDs, as well as the distance at which multiple markers on a single \ac{UAV} will start merging and mixing together their signals due to image proximity.
\subsection{Optical attacks}
\label{sec:discussion:attack}
The optical receiver can be attacked by flashing multiple \ac{UV} lights to the cameras. To understand how this attack operates let us first remember that each time a bright spot is detected the image processing module  assigns to it a serialized service number, it starts to track it, estimates its location, creates a dedicated buffer to store the associated samples and also creates a dedicated instance of classifier to process it. As mentioned before when the detector determines that such bright source does not correspond to a sequence in the dictionary $\mathbf{D}$ then all the this time series is killed and all the resources associated to it are liberated.

Therefore, if the attacker manages to generate a very large number of bright source spread over a large area then the image processing module could become saturated and the optical receiver become unoperational for the duration of the attack. Nevertheless, note that deploying such attack requires a large number of flashing lights witin the \ac{FoV} of the flying \ac{UAV} which can be costly and complicated to achieve. Also note that if the bright lights of the attacker are on the ground then the \acp{UAV} could just increase their altitude in order to take the malicious bright spots out of their \ac{FoV} which would then neutralize the attack.  
}

\section{Conclusion}
\label{sec:conc}
In this paper, we studied the theoretical and practical aspects of \ac{UVDAR}: a camera-based optical identification system for \acp{UAV}. Herein, it was shown how to optimize the optical signals emitted by the \acp{UAV} in order to maximize the number of detectable \acp{UAV} while minimizing identification time. Through theoretical analysis and experiments, we demonstrated  that clock signal mismatches impose important limitations on the capacity of this visual identification system. This visual identification system was tested both indoors and outdoors, demonstrating successful operation with sufficient performance. 
The results of this work can be used to further optimize visual-based localization and identification systems, such as UVDAR, as well as to evaluate the capacity of this system as the base for an optical communication network for \acp{UAV}.

\section*{Declarations}

\begin{itemize} 
\item Ethical statements. Not applicable.
\item Authors' contributions: D.B.L. developed the main theory presented in the manuscript, participated in the writing of the main manuscript and performed the simulations. V.W. contributed to the development of the theory, participated in the writing of the main manuscript, performed the experiments and prepared the figures. M.S. and M.G. contributed with the design of the manuscript and the writing process.
\item Acknowledgements. Not applicable.
\item Funding. This work was partially funded by EU under ROBOPROX (reg. no. CZ.02.01.01/00/22\_008/0004590), by the Czech Science Foundation (GAČR) project no. 23-07517S, and by the CTU grant no. SGS23/177/OHK3/3T/13.
\item Data availability: Not applicable.

\end{itemize}
\bibliographystyle{bst/sn-mathphys}

\bibliography{ref2}

\end{document}